\documentclass[prl,reprint,showpacs]{revtex4-1}

\usepackage{hyperref}
\usepackage{amsmath}
\usepackage{amsfonts}
\usepackage{amssymb}
\usepackage{epsfig}
\usepackage{graphicx}

\begin{document}

\title{Faraday rotation due to surface states in the topological insulator (Bi$_{1-x}$Sb$_{x}$)$_{2}$Te$_{3}$}

\author{Y. M. Shao$^{1,2,\ast}$, K. W. Post,$^{2}$, J. S. Wu$^{2}$, S. Dai$^{2}$, A. J. Frenzel$^{2}$, A. R. Richardella$^{3}$, J. S. Lee$^{3}$, 
\\N. Samarth$^{3}$, M. M. Fogler$^{2}$, A. V. Balatsky$^{4,5}$, D. E. Kharzeev$^{6,7}$ and D. N. Basov$^{1,2}$}

\affiliation
{$^{1}$Department of Physics, Columbia University, New York, New York 10027, United States\\
$^{2}$Physics~Department,~University~of~California-San~Diego,~La~Jolla,~California~92093,~United States \\
$^{3}$Department of Physics, The Pennsylvania State University, University Park, Pennsylvania 16802, United States\\
$^{4}$Nordita, KTH Royal Institute of Technology and Stockholm University, Roslagstullsbacken 23, SE-106 91 Stockholm, Sweden\\
$^{5}$Institute for Materials Science, Los Alamos National Laboratory, Los Alamos, New Mexico 87545, United States\\
$^{6}$Department of Physics and Astronomy, Stony Brook University, Stony Brook, New York 11794-3800, United States\\
$^{7}$Department of Physics and RIKEN-BNL Research Center, Brookhaven National Laboratory, Upton, New York 11973, United States
}

\begin{abstract}
Using magneto-infrared spectroscopy, we have explored the charge dynamics of (Bi,Sb)$_2$Te$_3$ thin films on InP substrates. From the magneto-transmission data we extracted three distinct cyclotron resonance (CR) energies that are all apparent in the broad band Faraday rotation (FR) spectra. This comprehensive FR-CR data set has allowed us to isolate the response of the bulk states from the intrinsic surface states associated with both the top and bottom surfaces of the film. The FR data uncovered that electron- and hole-type Dirac fermions reside on opposite surfaces of our films, which paves the way for observing many exotic quantum phenomena in topological insulators.
\end{abstract}
\maketitle
Three-dimensional (3D) topological insulators (TIs) are insulating in the bulk but host gapless, topologically protected surface states (TSSs) \cite{hasan2010,qi2011}. The Dirac-like TSSs are robust against disorder, as they are protected by topological properties of the bulk electronic wave functions \cite{moore2010}. The linear band dispersion of TSSs \cite{hsieh2009,kong2011,zhang2011} has been identified by angle resolved photoemission spectroscopy (ARPES), whereas scanning tunneling spectroscopy (STS) \cite{cheng2010,jiang2012} has verified the distinctive $\sqrt{B}$ dispersing Landau levels (LLs) of the TSS. With the advent of the new generation of thin films of 3D-TIs, it became possible to minimize bulk conduction, yielding an electrodynamic response that is dominated by the TSS \cite{post2015,wu2015,dipietro2012}. However, when surface sensitive probes (ARPES, STS) are applied to these new TI films, they are only able to characterize the top, vacuum-facing surface of the sample. While transport measurements probe all conducting channels in TI thin films, the decomposition of the response to each individual TSS/bulk layer is not trivial.

Here we report on Faraday rotation (FR) and cyclotron resonances (CR) magneto-transmission spectroscopy of (Bi$_{1-x}$Sb$_{x}$)$_{2}$Te$_{3}$ (BST) thin films. The combination of these techniques yields a comprehensive characterization of the TSS dynamics both at the top and bottom surfaces, complementary to transport and ARPES data. \cite{kong2011,zhang2011,he2012,lin2013,he2013}. We stress that FR, the rotation of light polarization after passing through a medium in magnetic field, is particularly informative in distinguishing multiple conductive channels in complex materials \cite{basov2005,levallois2015}. Whereas both CR and FR experiments are sensitive to transitions between the LLs, only the FR allows one to unambiguously discriminate the response of electrons and holes via the sign of the Faraday angle $\theta_F$ \cite{crassee2011a,crassee2011}. Additional insights come from the magnetic field dependence of the LL transitions energy ($\Delta E_{n}$), satisfying the dipole selection rule \cite{jiang2007,li2013}: $\Delta|n|=\pm1$. For instance, in a system with parabolic bands, including bulk bands in TIs, LL transitions scale linearly with magnetic field as $eB/m$, where $m$ is the band mass (classical CR). Drastically different LL energies are anticipated and observed for Dirac fermions in TSSs \cite{gusynin2007a,jiang2007,li2013}:
\begin{equation}
E_{\pm n} = E_{DP} \pm v_F\sqrt{2e\hbar|n|B}
\end{equation}
where $E_{DP}$ is the Dirac Point energy, $n$ is the LL index, $v_F$ is the Fermi velocity and $B$ is the magnetic field. Therefore, TSS/bulk carriers can be further distinguished by their dispersion of LL transition energies with magnetic field ($\sqrt{B}$/linear in $B$). 

\begin{figure*}[ht]
\centering
\includegraphics[width=177mm]{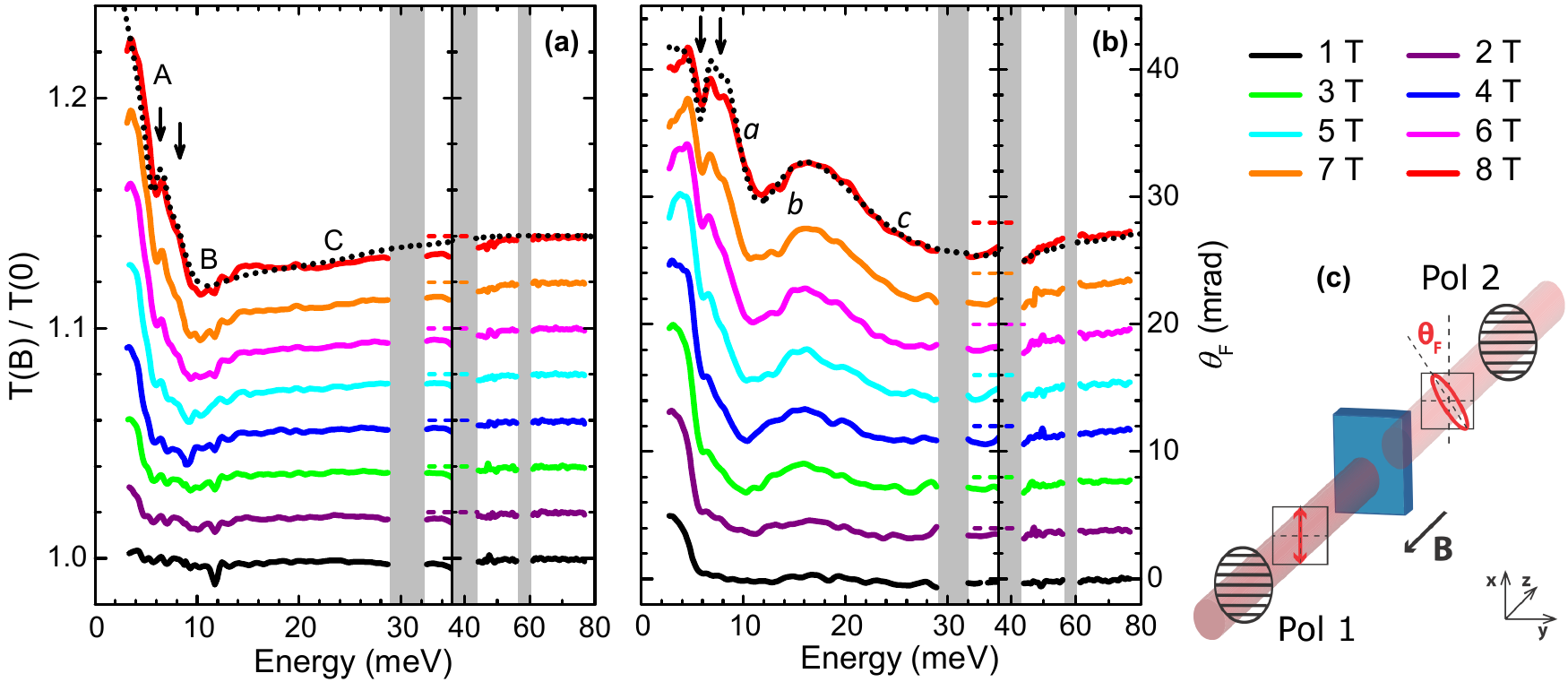}
\caption{Panel (a): magneto-transmission data normalized by the spectrum at zero field of the BST thin film. Panel (b):  Faraday rotation (FR) data. Data at different fields are offsetted by (a) 0.02 and (b) 4 mrad for clarity. Opaque regions due to InP absorption are greyed-out. Black dotted lines are Drude-Lorentz model fits for 8 T data and black arrows indicate the positions of infrared active phonons ($\sim$6 and 8 meV). Panel (c) is a schematic showing the FR definition and measurement setup.} 
\label{Figure 1}
\end{figure*}

Recently, we used terahertz/far-infrared transmission spectroscopy to study the dynamic response of BST thin films grown on InP (111)A substrates (see Supporting Information Sec. I for more details). We found an exceptionally small Drude weight that lies within the upper limit imposed by the conductivity sum rule for TSS \cite{post2015}. Although the sum rule criterion is instructive, this analysis does not allow one to discriminate between the contributions due to the surface and the bulk. In this work, we disentangle these contributions based on their unique, dynamic responses in a magnetic field. The broadband FR-CR measurements enable a multi-component analysis of the data and uncover marked distinctions between top and bottom TSSs in our films. The two surfaces have different Fermi energies and host Dirac fermions of opposite sign, which is required for an observation of a number of exotic effects, including topological exciton condensation in TIs \cite{seradjeh2009,zhang2011}. 

Generally, multiple types of carriers (electrons/holes in TSS/bulk bands) can coexist in TIs and all contribute to the magneto-optical (MO) response. According to the Drude-Lorentz (DL) model, the total conductivity is the sum of oscillators:
\begin{equation}
\begin{aligned} 
\sigma_{xx}(\omega,B) 
&=\sum_{j=1}^{N}\frac{D_j}{\pi}\frac{\gamma_j-i\omega}{\omega_{c,j}^2-(\omega+i\gamma_j)^2}\\
&-\sum_{j=1}^{N}\frac{i\omega\omega_{p,j}^2\varepsilon_{0}}{\omega_{o,j}^2-\omega^2-i\omega\Gamma_j}-i\omega\varepsilon_{0}(\varepsilon_{\infty}-1)
\end{aligned}
\end{equation}

\begin{equation}
\begin{aligned}
\sigma_{xy}(\omega,B)
&=\sum_{j=1}^{N}\frac{D_j}{\pi}\frac{-\omega_{c,j}}{\omega_{cj}^2-(\omega+i\omega\gamma_j)^2} 
\end{aligned}
\end{equation}

The first term in (2) and (3) stands for the Drude response of free carriers in magnetic field. Each oscillator is characterized by its own Drude weight ($D_j=\pi ne^2/m$), cyclotron frequency $\omega_{c,j}=qB/m$ and scattering rate $\gamma_j$. The second term in (2) represents Lorentzian oscillators centered at $\omega_{o,j}$, with plasma frequency $\omega_{p,j}$ and scattering rate $\Gamma_j$. The Lorentzian terms describe the contributions of phonons and are absent in the Hall conductivity $\sigma_{xy}$. The high frequency dielectric constant $\varepsilon_{\infty}$ accounts for absorptions above the measured frequency range, and is associated with interband transitions. In Eq. (2), each of the Drude oscillators can decribe a CR/LL transition in the classical/quantum regime. The model implies that the Drude absorption peak is shifted from zero frequency at $B=0$ to finite cyclotron frequency $\omega_c$ in the presence of an external magnetic field. Similar multi-component models have been used to successfully decode the complex MO spectrum for multilayer graphene \cite{crassee2011a} and Bi$_2$Se$_3$ thin films \cite{wu2015}. 

In the far-infrared region under the thin-film approximation, the measured broadband CR and FR spectra are directly related to real part of $\sigma_{xx}(\omega,B)$ and $\sigma_{xy}(\omega,B)$, respectively. For simplicity, $\sigma_{xx}$ and $\sigma_{xy}$ will be assumed to be the real part of the corresponding complex functions in the rest of the paper.

Magneto-transmission spectra normalized by the spectrum at zero field T($\omega,B$)/T($\omega,0$) are shown in Fig. 1(a). The most prominent field-induced features can be identified in the spectrum obtained at $B$ = 8 T: a marked increase ($\sim$10$\%$, labeled as A) in the transmission is observed at low frequency, followed by a dip (B) and a weak downturn (C) at higher frequencies. The gross features of the experimental data are accounted for within the Drude model of the cyclotron resonance (Eq. (2)). Both A and B features can be identified in the data at smaller fields. As the magnetic field increases (from 2 T to 8 T), the positions of these features increase: an observation consistent with the multi-component CR interpretation of the observed features.  The black arrows in Fig. 1 indicate the position of two phonon modes ($\sim$6 and 8 meV) in BST \cite{post2015}. The phonon frequencies do not exhibit frequency shifts with magnetic field.

\begin{figure}[ht]
\includegraphics[width=84mm,keepaspectratio=true]{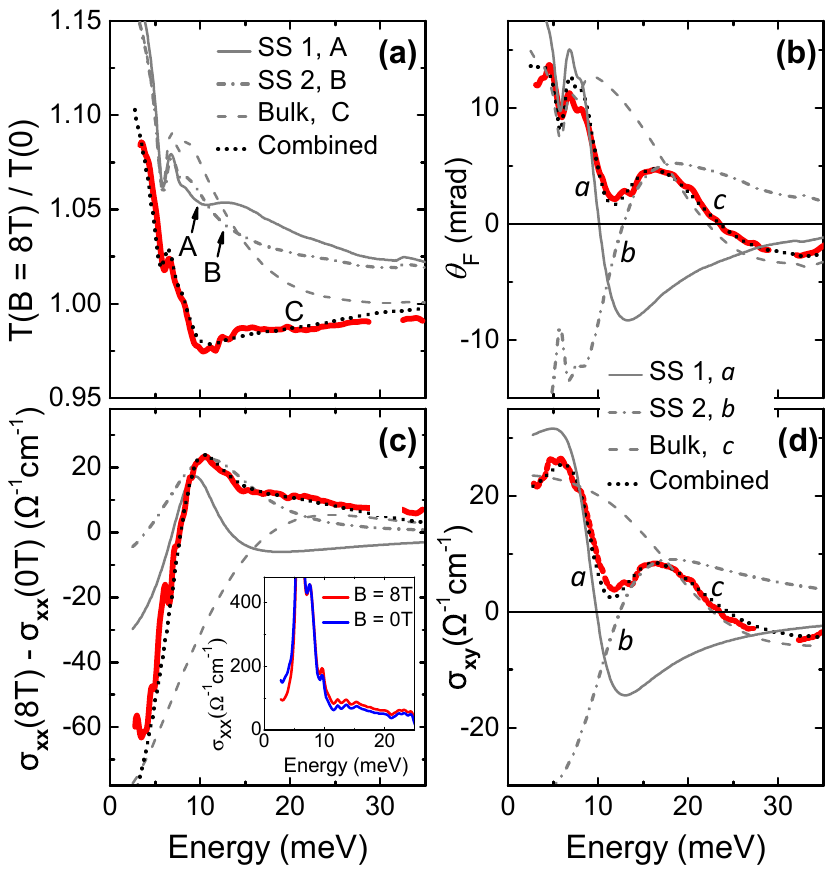}
\caption{Experimental and modeling results for (a) transmission and (b) FR spectra at 8 T using multi-component Drude-Lorentz model. In each panel, the red solid line is the data and black dotted line is the fit. Dashed grey lines represent contributions from the bulk. Solid and dash-dotted grey lines represent contributions from the two surfaces states. In panel (c) we plot the difference between the diagonal conductivity ($\sigma_{xx}$) spectra obtained at 8 T and at 0 T. The inset displays $\sigma_{xx}$ at 8 T and 0 T, which are dominated by phonons (6 and 8 meV) of BST. In panel (d) the Hall conductivity ($\sigma_{xy}$) is plotted. In (b) and (d) the signal from SS 2 (n-type) is opposite in sign compared to SS 1 and bulk (p-type).}
\label{Figure 2}
\end{figure}

Faraday rotation spectra, shown in Fig. 1(b), offer complementary insights into the magneto-optical response. In addition to the low-frequency phonons (arrows), one can readily observe three inflection points (\textit{a}, \textit{b} and \textit{c}) below 30 meV. Each inflection point corresponds to a peak in $\sigma_{xx}$ associated with an underlying CR/LL transition \cite{crassee2011a,wu2015}. All three inflection points harden with increasing magnetic field.

The highest measured FR angle in the low-frequency limit of our data ($\sim$15 mrad) is an order of magnitude smaller compared to previous work on Bi$_2$Se$_3$ \cite{wu2015}. We remark that the quantized FR is expected to occur in multiples of $\alpha\sim7$ mrad both in graphene and TIs in the limit of $\omega\ll\omega_c,E_g$ \cite{tse2011,morimoto2009}, where $E_g$ is the energy gap associated with the bulk response. The experimental observation of quantized FR has been reported recently in different TI systems at THz frequencies \cite{shuvaev2016,okada2016a,wu2016}. In this work, the condition $\omega\ll\omega_c$ is not strictly fulfilled since the lowest experimental frequency ($\sim$3 meV) is still very close to $\omega_c$ ($\sim$8 meV) at 8 T due to the broadness of the resonance. Therefore, the observed small rotation is unlikely to be related to quantized FR. Below we will demonstrate that the small rotation is the result of an interplay of multiple transitions, clearly resolved in our broadband spectra, yielding both negative and positive FRs. 

We used the DL model (Eqs. (2),(3)) to fit the CR and FR spectra at each measured magnetic field. The minimum model required to fit both data set contains three different Drude oscillators. A plausible interpretation of these three oscillators is in terms of three conduction channels associated with the bulk, top and bottom surfaces of our films. In Figs. 2(a), 2(b) we plot each of these three contributions along with the experimental data at 8 T (red solid line).

The coexistence of three mobile carriers in BST is particularly evident from the decomposition of the fit for $\theta_{F}$ (Fig. 2(b)). The gross features of the spectra are well described by the multi-component FR response \cite{crassee2011a}. Specifically, the two features with negative slopes (\textit{a} and \textit{c}) attest to two distinct p-type carriers with different Drude weight and scattering rate. We depict these contributions with the solid and dashed grey lines, respectively. The hump feature with positive slope (\textit{b}) originates from n-type carriers and is indicated with the grey dash-dotted line. In the limit of small Faraday rotation ($\theta_{F}\ll$ 1 rad) \cite{shimano2013}, the total rotation angle equals the sum of rotations due to the three terms \textit{a}, \textit{b} and \textit{c}. This net rotation (black dotted line) reproduced all features of our data (red line). The very same parameter set also captures the behavior of magneto-transmission spectra in Fig. 2(a). All fitting parameters for different magnetic fields are summarized in Fig. 3 (see fits for all magnetic fields in Supporting Information Fig. S1). We stress that the same model describes the entire data of CR and FR spectra at all different fields. This consistent description of the entire data set attests to the validity of the multi-component analysis. 

The extracted diagonal conductivity $\sigma_{xx}$ at $B=0$ T and $8$ T are plotted in the inset of Fig. 2(c). As mentioned before, two phonons dominate the low energy conductivity and do not change with the magnetic field. Their frequencies ($\omega_o$), oscillator strengths ($\omega^2_{p}$) and widths ($\Gamma$) are therefore kept constant at their zero field values (listed in Table S1 in Supporting Information) during the fitting at all magnetic fields. The electronic background is modified by the field. Specifically, the spectral weight (SW), defined as SW=$\int_{0}^{\Omega_c}\sigma_{xx}d\omega$ ($\Omega_c$: cutoff frequency), shifts from low energy ($\sim$3 meV) to higher energy ($\sim$15 meV) in the $B=8$ T spectrum. This SW shift is more evident in the field-induced changes of $\sigma_{xx}$ (Fig. 2(c)). According to the $f$-sum rule \cite{basov2011}, the total SW should be conserved and independent of magnetic field, given an appropriate $\Omega_c$. Indeed, from Fig. 2(c) we find that the magnetic field induced accumulation of SW around $\hbar\omega_c$ is qualitatively compensated by loss of SW at low energies. 

Phonon absorptions also affect the shape of FR spectra at low frequencies, shown in Fig. 2(b). On the other hand, according to Eq. (3), the Hall conductivity ($\sigma_{xy}$) is expected to only contain electronic contributions. Indeed, the $\sigma_{xy}$ spectra (red line in Fig. 2(d)) extracted via thin-film approximation are free of phonon structures. The extracted total $\sigma_{xy}$ spectra show an excellent agreement with the three-layer DL model calculation (black dotted line), justifying the use of thin-film approximations (see Supporting Information Sec. I). 
\begin{figure}[!h]
\includegraphics[width=84mm,keepaspectratio=true]{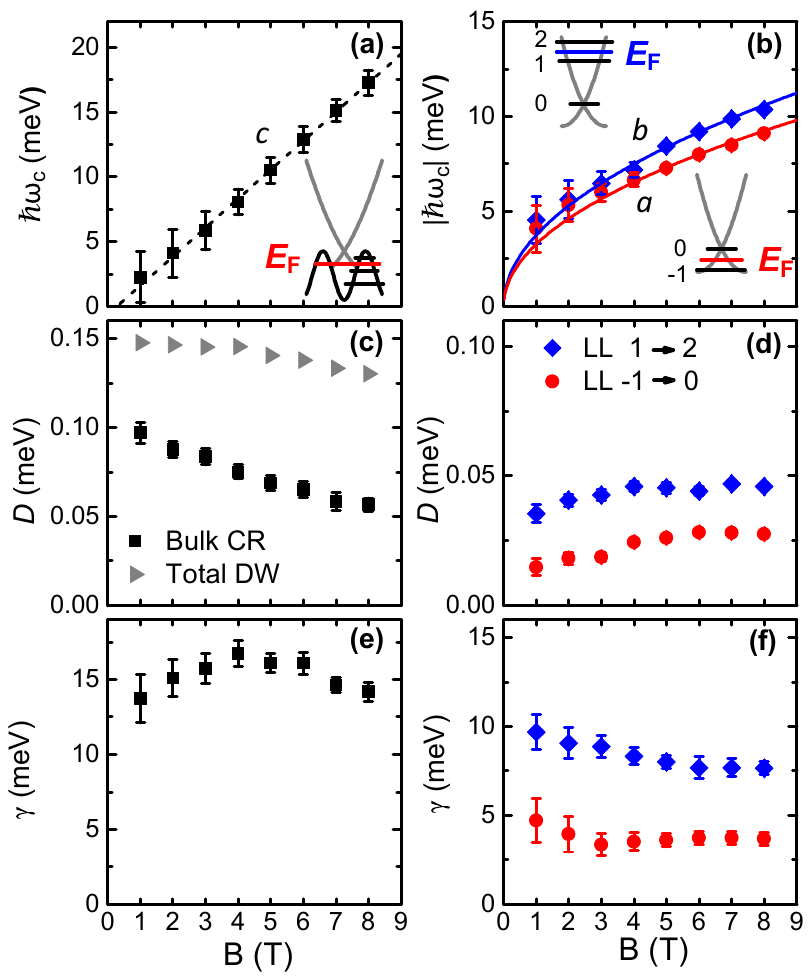}
\caption{Parameters obtained from the Drude-Lorentz fit of magneto-transmission and Faraday rotation spectra (Fig. 1). (a) Transition energies of the CR-like peak for bulk carriers; dashed line is a linear fit to the data. (b) Intraband LL transition energies for SS 1 (red circle) and SS 2 (blue diamond). Solid lines are model curves (Eq. (1)) of LL$_{1\rightarrow2}$ and LL$_{-1\rightarrow0}$ with $v_{F}=2.49\times10^5$ m/s and $0.9\times10^5$ m/s, respectively. Drude weight of the bulk and TSS carriers are displayed in (c) and (d), respectively. The decreasing Drude weight of the bulk CR is unexpected for typical CR transitions. The sum of all three Drude weights is also displayed in (c) as grey triangles. (e) and (f) are scattering rates for bulk and TSS carriers.}
\label{Figure 3}
\end{figure}
We now turn to the analysis of the fitting parameters extracted from the three-layer DL description of magneto-optics data. Each layer is parameterized by its own cyclotron frequency ($\hbar\omega_c$), Drude weight ($D$) and scattering rate ($\gamma$). All these contributions are additive in the MO response in the thin-film limit \cite{wu2015,shimano2013}.

In Fig. 3 we plot the field dependence of $\hbar\omega_c$, $D$ and $\gamma$ extracted from simultaneous fitting of the CR and FR spectra at different magnetic fields. This analysis reveals the $\sqrt{B}$-dependence of the transition energy for the two lower energy transitions, shown in Fig. 3(b). The $\sqrt{B}$ law is the hallmark of LL spacings for massless Dirac fermions and therefore we assign these two terms to the response of the TSSs in our films. Notably, this behavior can only be observed provided the Fermi energy is close to the Dirac point when only the lowest few LLs are occupied \cite{crassee2011,orlita2012}. We thus conclude that the Fermi energies associated with two surfaces in BST are very close to the Dirac point and we will report estimates of $E_F$ below. In contrast, the broader CR-like transition (Fig. 3(a)) shows conventional linear field dependence. We attribute this latter mode to bulk p-type carriers, as pictured in the inset of Fig. 3(a). The cyclotron mass extracted from the slope is 0.055 $m_e$, which is smaller than the effective mass of 0.08 $m_e$ deduced from quamtum oscillations for a p-type BST sample \cite{kohler1977}. The discrepancy could be due to difference in Fermi energy (BST valence band are non-parabolic \cite{kohler1977}) and/or stain in our thin films. Note that there is a peak structure in $\gamma$ (Fig. 3(e)) around 4 T, where the corresponding $\hbar\omega_{c}\sim8$ meV coincides with one of the phonons of BST. It is possible that the peak arises from the scattering between bulk CR and phonon modes, as suggested in a recent terahertz MO study on Bi$_2$Se$_3$ \cite{wu2015}. 

It is worth emphasising that the transition energies of the two SS transitions (a and b) are fairly close to each other, especially at low fields. We can nevertheless differentiate these two contributions because of the hump structure in the FR spectra, which is characteristic of the two opposite polarization rotations nearly canceling each other. Thus, the FR spectra allow us to unambiguously differentiate and assign these two responses to two different surfaces.

The evolution of Drude weight and scattering rate of the two SS transitions with field are intriguing as well (Figs. 3(d), 3(f)). Both Drude weights increase with increasing magnetic field, in accord with the theoretical prediction for Dirac fermions \cite{gusynin2007a}. In contrast, the Drude weight for semiclasical CR is usually field-independent and equals the zero-field Drude weight \cite{orlita2012}. The scattering rates of the two SS transitions decrease with $B$ field and show an overall $1/B$ dependence, which is indicative of charged impurity scattering \cite{yang2010,orlita2011}. The scattering rates of the TSS carriers are also much smaller than their bulk counterpart, indicating much higher mobility associated with the surface states.

The requirement of $E_F$ lying very close to the Dirac point ($\sqrt{B}$ dispersion) constrains the assignment of the observed SS Landau level transitions to two possibilities: LL$_{0\rightarrow1}$ and LL$_{1\rightarrow2}$ or alternatively LL$_{-1\rightarrow0}$ and LL$_{-2\rightarrow-1}$ for hole-like transitions. Higher intraband LL transitions ($|n|\geqslant2$) are excluded for two reasons: (i) At $|n|\sim2$, these transitions are expected to produce discontinuous jumps in transition energy that are not observed in our data; (ii) At $|n|\gg2$, corresponding to high Fermi energy, the evolution of the transition energy with the B field resembles the linear scaling (quasi-classical CR)\cite{witowski2010}, whereas our data show the $\sqrt{B}$ behavior, requiring low Fermi energy. Fitting the two SS transitions with linear dependence will result in finite intercept in energy, contrary to the semiclassical CR picture. On the other hand, fits with $\sqrt{B}$ dependence show excellent agreement with experimental transitions energies (Fig. 3(b)). Since the n-type SS transition also has higher Drude weight than the p-type SS, we attribute the former to LL$_{1\rightarrow2}$ and the latter to LL$_{-1\rightarrow0}$, with $E_F\sim25$ meV and $\sim-6$ meV away from the Dirac point, respectively (see Supporting Information Figs. S3, S4). 

The $\sqrt{B}$ fitting in Fig. 3(b) assumes a Fermi velocity of $2.49\times10^5$ m/s ($0.9\times10^5$ m/s) for the n-type (p-type) TSS. The $v_{F}$ of our n-type TSS is very close to the ARPES value of $2.2\times10^5$ m/s on an n-type Sb$_2$Te$_3$/Bi$_2$Te$_3$ bilayer film with similar $E_F$ ($\sim$ 30 meV) \cite{eschbach2015}. Due to the electron-hole asymmetry, $v_{F}$ is expected to be smaller on the hole side of the Dirac cone, consistent with our smaller estimated $v_F$ for the p-type TSS.

Finally, we wish to remark on possible relevance of our findings to the hypothesis of chiral magnetic effect \cite{kharzeev2010,qi2008}. It was conjectured that TSSs separating bulk from vacuum become oppositely charged in applied magnetic field, qualitatively consistent with our data. The carrier density on the two surfaces was predicted to scale linearly with the magnetic field, also in approximate agreement with our data (Fig. 3(d)) at $B<$ 5 T. The induced densities flatten off at $B>$ 5 T, likely due to the screening of finite bulk charge carriers. This hypothesis has implications for the finite frequency response of a TI: a resonance mode can be anticipated in infrared frequencies, offering yet another opportunity to investigate the condensed matter manifestations of phenomena discussed in high energy physics. The search for such resonances remains a challenge for future theoretical and experimental studies of 3D-TIs as well as Dirac/Weyl semimetals \cite{kargarian2015,akrap2016} in magnetic field. 

In conclusion, we optically distinguished coexisting TSSs and bulk carriers in BST from their different LL transition dynamics in magnetic field. Our results also show that the Fermi levels for the TSSs are remarkably low, allowing for the observation of $\sqrt{B}$ behavior characteristic of intraband LL transition. Importantly, the top and bottom TSS host carriers of opposite type, separated by the bulk. Such separated n- and p-type Dirac fermions paves the way for the observation of exotic quantum phenomena in TI, such as topological magneto-electric effect \cite{morimoto2015a} and topological exciton condensation \cite{seradjeh2009}.

\textit{Acknowledgements} - This works is supported by DOE Grant NO. DE-FG02-00ER45799. Sample growth and characterization at Penn State was supported by ONR (Grant No. N00014-15-1-2370)
and ARO-MURI (Grant No. W911NF-12-1-0461). D.K. is supported by DOE Grants No. DE-FG-88ER40388 and DE-SC0012704. D.N.B. is the Moore Foundation Investigator,
EPIQS Initiative Grant GBMF4533.

\bibliographystyle{apsrev4-1}

\bibliography{BST_MO}

\end{document}